\documentclass[fleqn,10pt,superscriptaddress]{wlscirep}

\usepackage{graphicx,amsfonts,amsmath}
\usepackage{dcolumn}
\usepackage{bm}
\usepackage{color}
\usepackage{subfigure}
\usepackage{esint}

\newcommand{\bk}{b_{\bm{k}}}
\newcommand{\bkd}{b_{\bm{k}}^{\dag}}
\newcommand{\bmk}{b_{-\bm{k}}}

\newcommand{\sumk}{\sum_{\bm{k}}}
\newcommand{\wk}{\omega_{\bm{k}}}
\newcommand{\fk}{f_{\bm{k}}}
\newcommand{\gk}{g_{\bm{k}}}
\newcommand{\pe}{|1\rangle\!\langle 1|}

\newcommand{\rl}{\rangle\!\langle}
\newcommand{\unit}{\mathbb{I}}
\DeclareMathOperator{\Tr}{Tr}

\title{Phonon-mediated generation of quantum correlations between quantum dot qubits.}

\author[1]{Jan Krzywda}
\author[1,*]{Katarzyna Roszak}

\affil[1]{Department of Theoretical Physics,
	Faculty of Fundamental Problems of Technology, Wroc{\l}aw University of Technology,
	50-370 Wroc{\l}aw, Poland}

\affil[*]{Correspondence to katarzyna.roszak@pwr.edu.pl}

\begin{abstract}
	We study the generation of quantum correlations between two excitonic quantum dot qubits due to their interaction with the same phonon environment.
	Such generation results from the fact that during the pure dephasing process
	at finite temperatures, each exciton becomes entangled with the phonon
	environment.
	We find that for a wide range of temperatures quantum
	correlations are created due to the interaction. The temperature-dependence
	of the level of correlations created displays a trade-off type behaviour;
	for small temperatures the phonon-induced distrubance of the qubit states
	is too small to lead to a distinct change of the two-qubit state, hence,
	the level of created correlations is small,
	while for large temperatures the pure dephasing is not accompanied
	by the creation of entanglement between the qubits and the environment,
	so the environment cannot mediate qubit-qubit quantum correlations.
\end{abstract}
\begin{document}
	
	\flushbottom
	\maketitle
	%
	%
	\thispagestyle{empty}

	\section*{Introduction}
	
	Quantum correlations play a crucial role in the understanding and possible implementation of any quantum computation algorithm \cite{Ekert_RMP96,shor97,grover97}. Unfortunately the influence of the environment is usually hostile to entanglement\cite{yu04,dodd04,roszak06a}, which is the standard type of quantum correlations used for quantum computation. 
	It has been recently shown that a weaker type of quantum correlations,
	those which are measured by the quantum discord\cite{olliver01,henderson01,modi12} and which are sometimes
	present in separable (nonentangled) states, are also useful from the perspective
	of quantum computation\cite{meurer92,knill98,datta08,passante11,horodecki14,
		dakic12,silberhorn13,fedrizzi13,peuntinger13,vollmer13}. Typically, an interaction with the environment
	is also detrimental to the quantum discord, but the discord is expected
	to be much more robust under the influence of the environment, and
	under some conditions may even be enhanced by noise\cite{modi12}.
	
	Quantum dot (QD) excitonic qubits\cite{zanardi98b}, for which one of the qubit states
	is an empty QD and the other is a ground state exciton
	excited in the QD, unavoidably interact with a bath of vibrations of the
	lattice of the crystal in which the dot is embedded (phonon
	environment)\cite{krummheuer02,vagov02a,borri01,vagov03,vagov04}. This interaction is diagonal in terms of QD states and and hence, can only lead
	to pure dephasing of excitonic 
	qubits.
	Contrarily to interactions with environments which are not diagonal
	in the subspace of qubit eigenstates, some of which have been shown 
	to lead to the generation of inter-qubit entanglement 
	\cite{mccutcheon09,cotlet14, benatti09}, the interaction does not lead
	to entanglement between the qubits.
	Yet it has been recently shown
	that such a process is, at finite temperatures, always accompanied by the creation of entanglement between the qubit and the environment\cite{roszak15a},
	and hence, we can expect some kind of quantum correlations to be 
	generated between qubits via the interaction with a common phonon environment .
	
	We study a system composed of two QD excitonic qubits
	separated by a finite distance and, hence, interacting with a common
	environment of phonons. We find that such an interaction
	will lead to the creation of finite quantum discord values between the
	two qubits, if the distance between them is small enough that the
	environments cannot be treated as separate,
	and the temperature is modest. Because of the characteristic ``partiality''
	of phonon-induced processes, the generated discord is robust 
	until the influence of other, slower decoherence mechanisms become
	dominant. We identify two most prominent features of the evolution
	during the generation of the quantum discord and study their origin
	and parameter dependence (which are both different)
	with the help of X-states whose quantum correlations are easier to quantify.

	
	\section*{Results}


	\subsection*{Generation of quantum correlations}
	
	The system under study consists of two QD excitonic qubits,
	where the $|0\rangle$ and $|1\rangle$ qubit states denote the empty dot 
	and the single exciton in its ground state, respectively.
	Such excitonic qubits are known to suffer from a strong interaction 
	with an environment of phonons (quanta of the vibrations of the crystal
	lattice in which the dots are embedded)\cite{krummheuer02,vagov02a}.
	The interaction leads to pure dephasing of the qubit states
	which is only partial (the qubit coherences decrease up to some finite
	value which is strongly dependent on temperature) due to the super-Ohmic nature of the phonon bath. As has been recently shown in Ref.~[\citenum{roszak15a}], the dephasing process is accompanied
	by the creation of entanglement between the qubit and its phonon environment.
	Since the two QDs interact with the same phonon environment
	(unless they are infinitely distant from each other)
	it is reasonable to expect a generation of quantum correlations 
	between the two qubits which would result
	from the fact that both of them become entangled with a common bath.
	
	\begin{figure}[ht]
		\centering
		\includegraphics[width=0.8\textwidth]{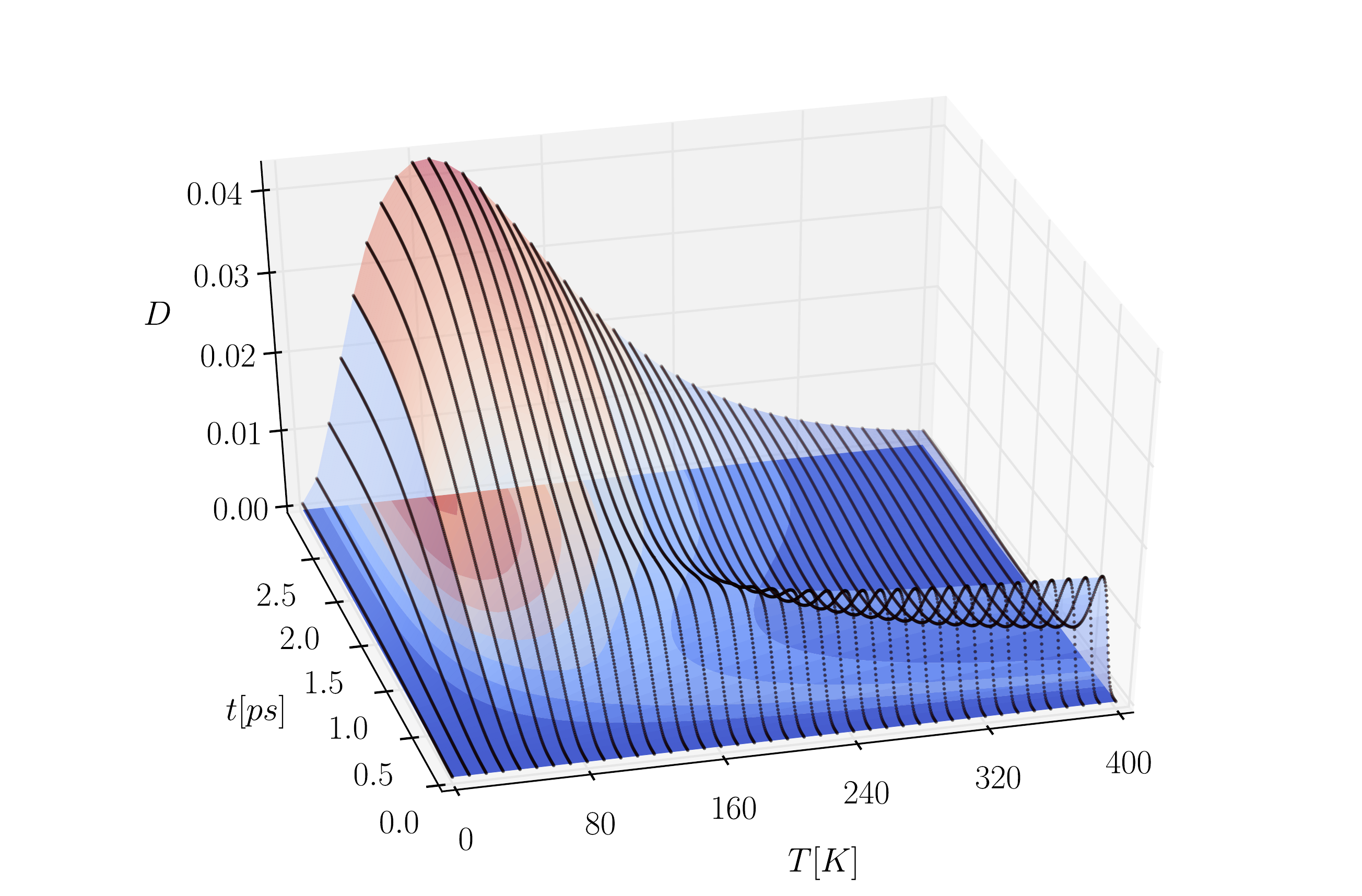}
		\caption{\label{fig:evol} Evolution of the rescaled discord as a function of time and temperature for the initial two-qubit state (\ref{ini2})
			with $\alpha=\beta=1/\sqrt{2}$
			at the distance between dots $d=6$ nm.}
	\end{figure}
	
	We study the evolution of quantum correlations between two qubits in a pure initial state which is not correlated in any way (quantumly or 
	classically).
	To this end we assume that both qubits are in the same pure initial state
	given by 
	$
	|\Psi\rangle = \alpha|0\rangle +\beta|1\rangle,$
	where $|\alpha|^2+|\beta|^2=1$.
	The initial two-qubit state is a product of the two single qubit
	states and is then given by
	\begin{equation}\label{ini2}
	|\Psi\rangle_L\otimes|\Psi\rangle_R = \alpha^2|0\rangle +\alpha\beta|1\rangle
	+\alpha\beta|2\rangle+\beta^2|3\rangle,
	\end{equation}
	where the indices $L$ and $R$ differentiate between the two dots on the
	left side of the equation and a two-qubit basis is used on the right side
	with
	$|0\rangle\equiv|0\rangle_L\otimes|0\rangle_R$,
	$|1\rangle\equiv|0\rangle_L\otimes|1\rangle_R$,
	$|2\rangle\equiv|1\rangle_L\otimes|0\rangle_R$, and
	$|3\rangle\equiv|1\rangle_L\otimes|1\rangle_R$.
	Obviously the state is separable and it also has zero discord.
	
	We also assume that the whole system, consisting of the two qubits and
	the phonon environment, is initially in a product state, and that the
	environment is then at thermal equilibrium.
	The Hamiltonian of this system and the Weyl operator method which 
	allows for the exact diagonalization of such Hamiltonians are presented
	in the Methods Section. The exact formulas which govern the evolution 
	of the two QDs after the degrees of freedom of the environment are traced
	out are also explicitly stated there together with a detailed discussion
	of the behavior of the two qubit density matrix in general and in the long-time, high-temperature, and large-distance between the dots limits. 
	
	In the following we do not take into account the interaction between the qubits
	(the biexcitonic shift which describes the energy shift of the state $|11\rangle$ due to the interaction when excitons are present in both dots). This is because the presence of the biexcitonic shift does not change
	the level of correlations generated between the qubits due to 
	the interaction with the environment, while it leads to 
	oscillations between the initial separable state and the maximally entangled state $1/2(|00\rangle+|01\rangle+|10\rangle-|11\rangle)$.
	In the presence of phonons these oscillations are damped
	while finite periods of time appear when entanglement is equal to zero
	\cite{roszak06a} (sudden death type behaviour followed by the rebirth
	of entanglement).
	In the case of the quantum discord, the same oscillations are seen, but 
	phonon dephasing leads to characteristic indifferentiable behavior
	near the state with highest entanglement, while discord curves are smooth
	around the zero-discord states \cite{roszak13}.
	
	It turns out that the described process can only lead to emergence
	of inter-qubit entanglement in extreme situations, for which the number of phonon modes
	taken into account is severely restricted and the exciton-phonon coupling 
	is extremely high. In all realistic situations, to which we limit ourselves
	here, the qubit-qubit interaction mediated by qubit-phonon entanglement
	can only lead to the appearance of weaker quantum correlations
	described by the quantum discord (while the two-qubit state remains 
	separable). Nevertheless, quantum correlations between qubits 
	do appear due to the
	process of phonon-mediated decoherence.
	
	The measure of quantum correlations used in the following is the rescaled
	discord\cite{tufarelli13}. This is a geometric measure which is related to the geometric
	discord. The geometric discord is defined as the smallest Hilbert-Schmidt distance between the studied state
	and the set of zero-discord states\cite{dakic10} and is one of the few discord
	measures which can be evaluated from the two qubit density matrix
	for any two qubit state.
	In fact, explicit formulas are known for the upper\cite{miranowicz12} and lower\cite{dakic10} bound
	on the geometric discord. The transition to the rescaled discord
	(which can be easily found for a pair of qubits, if the geometric discord is known)
	is made to ensure that the measure does not depend on the purity of the state
	(as the Hilbert-Schmidt distance does).
	The exact details needed for the calculation of the bounds on the geometric
	discord and the rescaled discord are given in the Methods Section.
	
	Fig.~\ref{fig:evol} shows the evolution of the rescaled discord
	(the lower and upper bounds of the discord coincide up to numerical
	precision for the data shown)
	of the initial state (\ref{ini2})
	with $\alpha=\beta=1/\sqrt{2}$ as a function of time and temperature
	with the distance between the dots is set at $d = 6$ nm
	(the studied dots lie on the same plane, are assumed identical, and are small and flat, with the height of the dots 
	set as $1$ nm and the width around $4$ nm; the width of the wavefuncitons
	describing the electron and hole are slightly different).
	In fact, the discord upper and lower bounds always coincide when $\alpha=\beta$;
	this is not the case when $\alpha\neq\beta$.
	As can be seen, the interaction with a common phonon environment 
	does lead to the appearance of quantum correlations between the two QDs.
	This is due to the fact that for any finite temperature the exciton-phonon interaction
	leads to entanglement between each qubit and the
	phonon environment as shown in Ref.~[\citenum{roszak15a}]
	(there is no time delay between the start of the joint exciton-phonon
	evolution and the appearance of entanglement). Since the dots are 
	coupled to the same environment (for any finite distance between them $d$),
	this allows for phonon-mediated transfer of quantum correlations.

	
	There are two prominent features of the generation of phonon-induced 
	quantum correlations which are both visible in Fig.~\ref{fig:evol}.
	The first is the appearance of a maximum at short times (sub-picosecond). 
	The second 
	is the accumulation of quantum correlations at slightly longer times
	(on the order of picoseconds)
	which leads to the emergence of a plateau. The effect of temperature on the
	plateau (in terms of its height and the time when it is reached) 
	is much stronger than the effect of temperature on the maximum
	and when the maximum is visible it survives for higher temperatures
	than the plateau.
	Both features are small at low temperatures, do not occur at infinite temperatures (since no exciton-phonon entanglement is created), and reach
	their maximal values at some finite temperatures (which are different
	for the two features).
	Contrarily,
	the effect of the distance between the dots is much stronger on the maximum
	and the maximum is visible only for very closely spaced QDs
	(the discord values at the maximum and at the plateau both grow
	with the inverse of the interdot distance).
	
\begin{figure}[ht]
		\centering
	\subfigure{
	\includegraphics[width=0.41\linewidth]{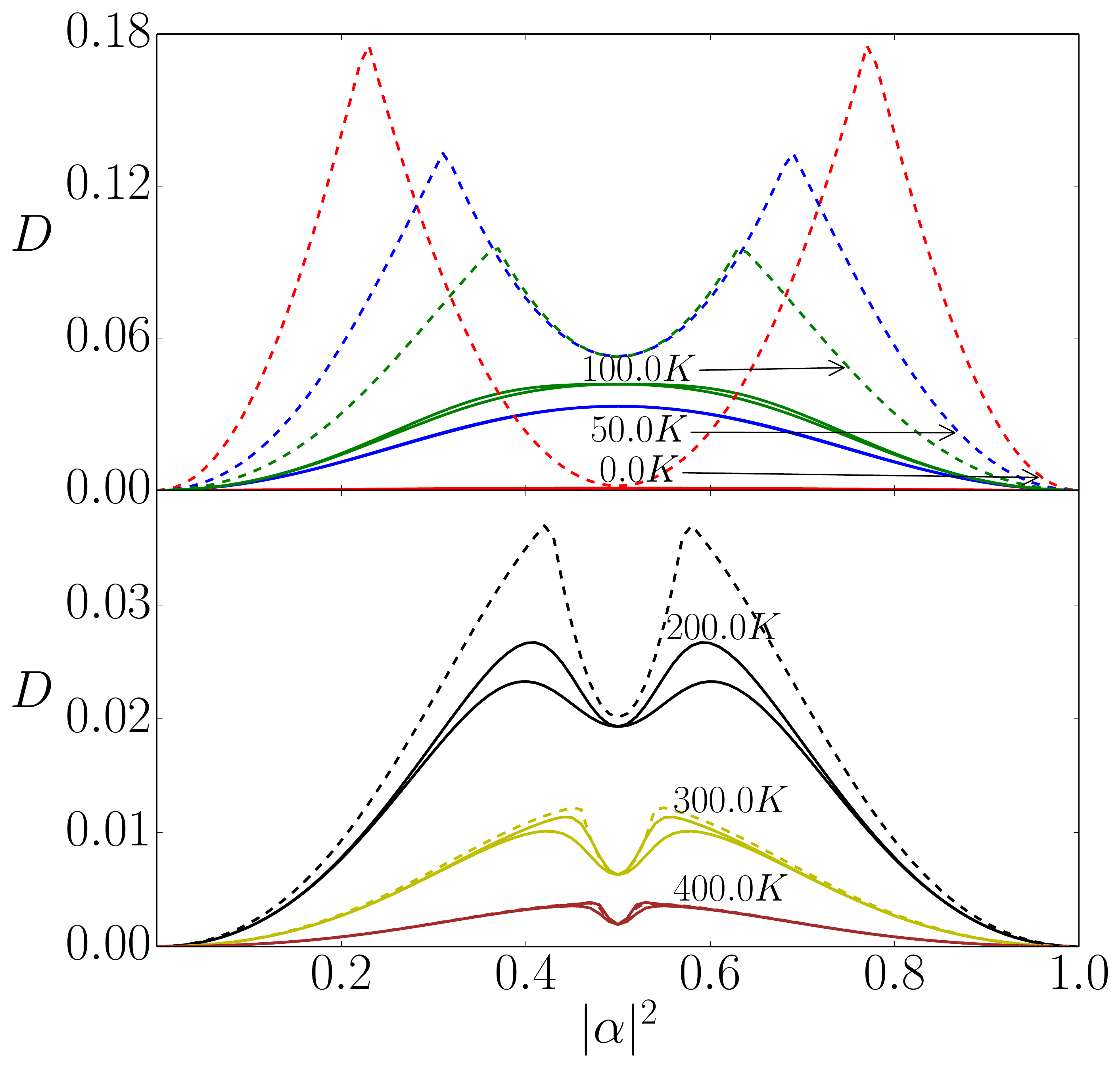}
}
\hspace{.5em}
\subfigure{
	\includegraphics[width=0.53\linewidth]{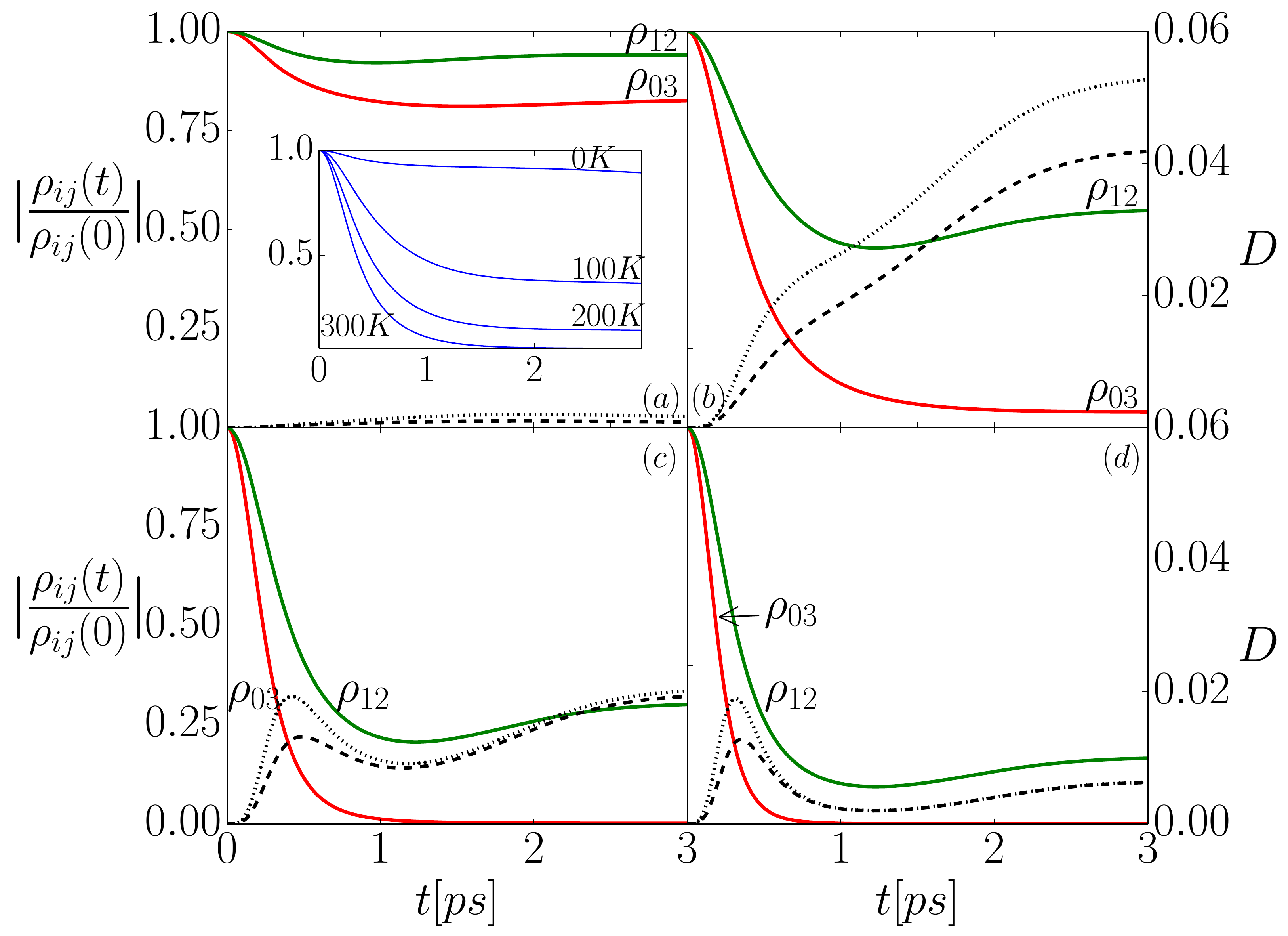}
}
		\caption{Left: Long-time (steady state) discord dependence on the single 
			qubit occupation $|\alpha|^2$ at different temperatures.
			Solid lines correspond to the pure initial state (\ref{ini2})
			and dashed lines correspond to the X initial state (\ref{x}).
			Top panel contains curves for $T\le 100$ K and bottom panel
			contains curves for $T> 100$ K.\\
			Right: 
			Evolution of normalized coherences at (a) 0 K,
			(b) 100 K, (c) 200 K, (d) 300 K. Red line - $\rho_{03}$, green line -
			$\rho_{12}$. Inset - $\rho_{01}$. The evolution of the rescaled
			discord corresponding to the coherences is given by the dashed black lines
			for initial pure state (\ref{ini2}) with $\alpha=\beta=1/\sqrt{2}$
			and by the dotted black lines for the respective initial X-state (\ref{x})
			with $a=b=c=x=y=1/4$.\label{fig2}}
	\end{figure}

	To understand the origin of the two features and the nature of their
	parameter dependences, it is necessary to understand what (phonon-induced) changes
	to the DQD density matrix cause them to emerge.
	To this end, the evolutions of the amplitudes of the normalized
	off-diagonal elements of the density matrix $|\rho_{ij}(t)|/|\rho_{ij}(0)|$
	are plotted in the right panel of Fig.~\ref{fig2} for different temperatures.
	The amplitude values corresponding to the elements $\rho_{03}(t)$ (describing
	the coherence between $|00\rangle$ and $|11\rangle$; red solid lines)
	and $\rho_{12}(t)$ (describing
	the coherence between $|01\rangle$ and $|10\rangle$; green solid lines)
	responsible for inter-qubit coherence
	are expected to be most important for the generation of quantum correlations.
	The amplitudes of all other coherences follow the same decay function
	which is plotted in the inset of the right panel of Fig.~\ref{fig2} (a).
	The decay of the three curves corresponding to different off-diagonal
	elements differs substantially. The decay of $\rho_{03}(t)$
	is faster than the decay of the other curves, since it is the coherence
	between states which are easiest to distinguish for phonons
	(the state when both dots are empty and the state when two excitons
	are excited). The coherences between states that differ by one exciton
	($\rho_{01}(t)$, $\rho_{02}(t)$, $\rho_{13}(t)$, and $\rho_{23}(t)$)
	evolve more slowly, but display the same type of decay as the $\rho_{03}(t)$
	curve. The element $\rho_{12}(t)$ describes the coherence between states
	which globally have one exciton which is either in the left or in the right
	dot,
	so they are the hardest for phonons to distinguish (especially for 
	small distances between the dots). Hence, the decay of this coherence qualitatively differs from all other
	coherences and it shows a slight revival at finite times which is due to the interference 
	of wave-packets from different QDs.

	The dashed black lines in the right panel of Fig.~\ref{fig2} show the evolution of 
	the rescaled discord for initial state (\ref{ini2})
	with $\alpha=\beta=1/\sqrt{2}$ (same as in Fig.~\ref{fig:evol}) corresponding to the
	decoherence curves shown in the same figure. Careful examination reveals that the initial
	maximum corresponds to the fast decay of the $\rho_{03}(t)$ curve
	with respect to the deacay of $\rho_{12}(t)$. This is 
	followed by a rise of the quantum discord until it reaches a plateau
	which corresponds to the revival of the $\rho_{12}(t)$ curve and
	it also reaching a plateau. 
	
	Further analysis of the processes which lead to the generation
	of quantum correlations here requires a simplified scenario
	for which the quantum discord can be calculated analytically.
	Since the situation under study involves only pure dephasing,
	an initial X-state will remain an X-state throughout the evolution.
	For an X-state, which is generally of the form
	\begin{equation}
	\label{x}
	\rho_X=\left(
	\begin{array}{cccc}
	a&0&0&y\\
	0&b&x&0\\
	0&x^*&b&0\\
	y^*&0&0&c
	\end{array}
	\right),
	\end{equation}
	the lower and upper bounds on the geometric discord coincide,
	and the value of the geometric discord can be easily found 
	using the formulas for the lower bound on the geometric discord
	given in the Methods Section. For 
	$1/2\left[(a-b)^2+(b-c)^2\right]\le(|y|+|x|)^2$ we get
	\begin{equation}
	\label{D2}
	D_S(t)= 2|y|^2 +2|x|^2,
	\end{equation}
	while for $1/2\left[(a-b)^2+(b-c)^2\right]<(|y|+|x|)^2$,
	the geometric discord is given by
	\begin{equation}
	\label{D1}
	D_S= 1/2\left[(a-b)^2+(b-c)^2\right]+(|y|-|x|)^2.
	\end{equation} 
	If $a=b=c=1/4$ and the initial X-state differs from the previously
	studied pure initial state only
	by four coherences which are set to zero,
	meaning that $a(0)=|\alpha|^4$, $b(0)=x(0)=|\alpha|^2|\beta|^2$, $c(0)=|\beta|^4$,
	and $y(0)=\alpha^2\beta^{*2}$, then the geometric discord is 
	in the regime described by Eq.~(\ref{D1}) at all times
	and is in fact equal to 
	$
	\label{Dx}
	D_S= (|y|-|x|)^2$.
	Obtaining the rescaled discord now only requires inserting this equation
	into Eq.~(\ref{rd}) as discussed in the Methods Section.
	
	The rescaled discord for the initial state (\ref{x})
	with $a=b=x=y=1/4$ (which resembles the pure, equal superposition state
	of Eq.~(\ref{ini2}) with $\alpha=\beta=1/\sqrt{2}$ the closest of all X-states)
	evolving in the same way as the equal superpositon state would is plotted in the right panel of Fig.~\ref{fig2}
	with black dotted lines. 
	The strong resemblance of the discord evolution of this X-state and 
	the corresponding pure state is striking. Including the presence and 
	the evolution of the four off-diagonal elements missing in the X-state
	changes only the quantitative features of the discord evolution
	(the discord generated in the pure state is slightly smaller),
	but the qualitative features remain the same. This means that the 
	appearance of the maximum and the plateau is of the same origin for the 
	two states and that it is the difference of the amplitudes of the two
	two-qubit coherences that is crucial for the phonon-induced generation
	of quantum correlations between qubits.
	Note, that the appearance of the maximum stems from the difference
	in the rates of the initial dephasing of the two off-diagonal elements
	and as such the process could not lead to the appearance or enhancement
	of entanglement. 
	This process does not in fact require generation of qubit-environment entanglement for the appearance of non-zero inter-qubit quantum discord to be possible \cite{modi12},
	which explains its much weaker temperature dependence than that
	of the plateau.
	The plateau, on the other hand, is related to the
	revival of the $\rho_{12}$ coherence, which is a common feature in
	few-qubit phonon decoherence, and is connected with the interference
	of phonon wave-packets travelling away from the dot. It can also lead to the revival of entanglement
	as seen in Ref.~[\citenum{roszak06a}].
	
	Finally, the study of the X-state resolves the problems of the dependence of the 
	generated quantum discord on temperature and distance between the dots.
	Firstly, for low temperatures, since there are few phonons in the system
	the resulting pure dephasing is weak, so the difference between the two
	coherences is small and so is the value of the quantum discord.
	For large temperatures, both coherences are strongly affected 
	(and the revival effect becomes negligible) which diminishes
	the difference between them, again leading to small values of the 
	quantum discord plateau (and to zero discord for infinite temperatures).
	Hence, the discord which is generated at the plateau is relatively high only for intermediate temperatures
	for which phonon effects are already strong while qubit-environment 
	entanglement is still effectively created.
	The maximum survives to higher temperatures, since it does not require
	the build-up of quantum correlations.
	As for the dependence on the distance between the quantum dots, the difference
	in evolution between $|\rho_{03}|$ and $|\rho_{12}|$ stems from the fact
	that both QDs interact with the same phonon environment. When the dots
	are infinitely far apart (separate phonon environments) the relation $|\rho_{03}|=|\rho_{12}|$
	is fulfilled at all times leading to no quantum correlations
	ever being generated.
	Both the difference between the rates of dephasing of $\rho_{03}$ and $\rho_{12}$ and the magnitude of the revival of $\rho_{12}$ decrease with
	growing distance between the dots 
	(the time of revival grows with this distance) while the environment with
	which they each interact becomes different.
	The plateau is much more robust against growing distance between the dots
	than the maximum, since in the initial stages of decoherence 
	(for short times) the evolution reaches the limit where it behaves
	similarly to the situation when each dot interacts with a separate environment
	for much smaller distances
	than in the later stages (for longer times). As distance grows, the plateau
	appears later in time (since it takes phonons from the two QDs longer
	to reach each other), but it does appear for all reasonable distances
	between the dots.
	
	
	
	
	Lastly, let us study the dependence of the generated
	quantum correlations on the initial single qubit state.
	To this end, long time (steady state) rescaled discord 
	of the initial state given by Eq.~(\ref{ini2}) is plotted as a function
	of single qubit occupation $|\alpha|^2$ (for $\alpha=|\alpha|$
	and $\beta=\sqrt{1-|\alpha|^2}$) in the plot on the left side of Fig.~\ref{fig2}
	(solid lines)
	at different temperatures.
	The first important feature is that for $\alpha\neq\beta$
	the lower and upper bounds on the discord do not necessarily coincide
	(both bounds are plotted in the figure).
	Secondly, for higher values of temperature (for which the dephasing
	is strong) the discord is not a convex function of $|\alpha|^2$
	(lower panel of the figure), 
	which could be expected and is true for lower temperatures
	(upper panel of the figure).

	
	The dependence of the long-time rescaled discord on $|\alpha|^2$ for the
	initial state (\ref{x}) with $a=|\alpha|^4$, $b=x=y=|\alpha|^2(1-|\alpha|^2)$
	and $c=(1-|\alpha|^2)^2$ is also shown in
	the plot on the left side of Fig.~\ref{fig2} (dashed lines).
	The comparison of
	the asymptotic discord of such states to that of the corresponding pure states is more tricky than in the case of
	$\alpha=1/\sqrt{2}$, because only for $\alpha=0$, $\alpha=1/\sqrt{2}$, 
	or $\alpha=1$ the discord of the initial state is equal to zero
	(so only then there are no quantum correlations present
	in the initial state). Otherwise the initial discord value is finite
	and the initial geometric discord is 
	obtained using Eqs (\ref{D2}) and (\ref{D1}), which yield
	\begin{equation}
	D_S = \min\left[
	\frac{1}{4}(2|\alpha|^2-1)^2\left[(2|\alpha|^2-1)^2+1\right],
	4|\alpha|^2(|\alpha|^2-1)^2
	\right].
	\end{equation}
	
	The long-time values of the X-state discord
	clearly show the transition between different regimes of decay
	(here as a function of $\alpha|^2$)
	which is characteristic of the discord\cite{Maziero_PRA09,mazzola10,roszak13,Lim_JPA14}.
	The transition is occurs when the dependence of the discord as a function of
	$|\alpha|^2$ changes between increasing and decreasing (on the left plots
	in Fig. \ref{fig2}).
	Note, that it occurs
	irrelevant of the temperature studied. 
	For low temperatures (the upper panel of the plot on the left side of Fig.~\ref{fig2})
	there is no resemblance in the $|\alpha|^2$-dependence between the
	pure-state (\ref{ini2}) and the X-state (\ref{x}). 
	This is because at low temperatures the four coherences which are initially
	zero for the X-state are preserved at reasonably large values in the pure state
	evolution. This is not the case for high temperatures
	and the level of long-time correlations present in the 
	initially pure two qubit system
	starts to resemble that of the X-state. Although there is no visible
	transition between discord decay regimes 
	for the initial pure states, the dependence on the single qubit occupation
	shows a similar pattern as that of the X-state.
	This is because for high temperatures the ``irrelevant'' coherences
	are quickly and largely reduced by the interaction with phonons
	and the pure state qualitatively resembles the X-state at long times.

	
	\section*{Discussion}
	
	We have studied the generation of quantum correlations between two QD
	qubits which do not interact directly, but are coupled to a common phonon
	bath. Such a qubit-environment interaction leads to the generation
	of entanglement at finite temperatures. As we have shown, this 
	suffices for the generation of quantum correlations described by
	the quantum discord (but not entanglement itself)
	between the qubits.
	Since phonon-induced decoherence is only partial,
	the long-time correlations generated are robust
	(until other slower decoherence sources, such as exciton recombination, 
	take over).
	
	The time-evolution of the two-qubit quantum discord
	of an initial pure product state displays two main features
	which depend in a non-trivial manner (and differently) on 
	temperature and interdot distance.
	The features result from different decoherence mechanisms
	and while the initial feature (maximum)
	could be caused by noise which does not result from the generation
	of qubit-environment entanglement
	the second feature (plateau) stems from the same process which leads
	to the slight rebirth of two-qubit coherence and 
	can reduce two-qubit entanglement decay.
	Hence, the plateau mechanism which is the one leading to robust
	discord generation requires entanglement between the qubits
	and their common environment.
	
	Finally, we have studied the dependence of the level of quantum correlations
	generated 
	on the initial single qubit occupations.
	This dependence becomes counterintuitive beyond some threshold
	temperature (around $150$ K for the studied system) and instead of 
	displaying monotonous behavior when the occupation changes from
	completely asymmetric ($|\alpha|^2=0$ or $|\alpha|^2=1$)
	to fully symmetric ($|\alpha|^2=1/2$),
	we observe a maximum at some value of $|\alpha|^2$ between zero 
	and one half and a local minimum is reached at $|\alpha|^2=1/2$.
	The origin of this behavior can only be understood with the help
	of corresponding X-states (which is reasonable at high temperatures), 
	the study of which reveals that the non-monotonicity is related
	to transitions between different decay regimes which is characteristic
	for quantum discord evolutions.
	
	\section*{Methods}
	\subsection*{Double quantum dot evolution in the presence of phonons}
	The system under study consists of two excitonic QD qubits located in one plane
	and interacting with a common phonon
	environment \cite{krummheuer02,vagov02a} The basis states of each qubit are $|0\rangle$ which corresponds
	to an empty QD and $|1\rangle$ which denotes an exciton in its ground state 
	excited in the QD. 
	The system is described by the Hamiltonian 
	$H=H_L+H_R+H_{ph}$,
	where 
	$H_i=\epsilon_{i}\pe+\pe\sumk\fk^{(i)}(\bkd+\bmk)$,
	with $i=L,R$ differentiating the dots, describes the energy
	of each dot and its interaction with the phonon environment.
	The third part of the Hamiltonian describes the free phonon energies
	$H_{ph}=\sumk\hbar\wk\bkd\bk$.
	Here, 
	$\epsilon_{i}$ are the transition energies
	in the two subsystems, $\fk^{(i)}$
	are system-environment coupling constants, $\bk,\bkd$ are bosonic
	operators of the phonon modes, and $\hbar \wk$ are the corresponding energies. 
	The energy shift
	due to the interaction between the subsystems (the biexcitonic shift)
	has been omitted here, since it leads to coherent two-qubit rotations
	and consequently to oscillations of entanglement that would only distort the 
	phonon-induced generation of quantum correlations that we want to 
	observe.

	This type of Hamiltonian can be diagonalized analytically 
	following Ref.~[\citenum{roszak06a}] by the
	transformation 
	$
	\mathbb{W}=|0\rl 0|\otimes\unit+\sum_{i=1}^{3}|i\rl i|\otimes w_{i},
	$ 
	where in the two-qubit basis the states denote
	$|0\rangle\equiv|0\rangle_L\otimes|0\rangle_R$,
	$|1\rangle\equiv|0\rangle_L\otimes|1\rangle_R$,
	$|2\rangle\equiv|1\rangle_L\otimes|0\rangle_R$, and
	$|3\rangle\equiv|1\rangle_L\otimes|1\rangle_R$.
	The operators $w_{i}$ are Weyl shift
	operators and are given by
	$
	w_{1,2}=\exp\big[\sumk\gk^{(1,2)*}\bk-\mathrm{H.c.}\big]$ and
	$w_{3}=\exp\big[\sumk(\gk^{(1)*}+\gk^{(2)*})\bk-\mathrm{H.c.}\big]$,
	where
	$\gk^{(i)}=\fk^{(i)}/\hbar\wk$. 
	The diagonalized Hamiltonian is given by
	$
	\label{hamd}
	\tilde{H}=\mathbb{W}H\mathbb{W^{\dag}}
	=E_{L}(|1\rangle_{LL}\langle 1|\otimes\unit_R)
	+E_{R}(\unit_L\otimes|1\rangle_{RR}\langle 1|)+H_{ph},
	$
	where the left side in the tensor product corresponds to the left dot
	($L$) and the right side to the right dot ($R$).
	The shifted energies here are given by
	$E_{i}=\epsilon_{i}-\sumk\hbar\wk|\gk^{(i)}|^{2}$.
	
	The evolution operator of the whole system
	may now be written in terms of the evolution operator of the 
	diagonalized Hamiltonian
	and Weyl operators as
	$U_{t}=\mathbb{W}^{\dag}\mathbb{W}_{t}\tilde{U}_{t}$, where
	$\tilde{U}_{t}=\exp(-i\tilde{H}t/\hbar)$ and
	$\mathbb{W}_{t}=\tilde{U}_{t}\mathbb{W}\tilde{U}_{t}^{\dag}$. 
	Hence, when the system and environment are initially
	in a product state,
	with the environment 
	in a thermal equilibrium state $\rho_{T}$, the reduced density matrix
	of the two-qubit system is given by
	$
	{\rho}(t)
	=  \mathrm{Tr_{R}}\left[
	\mathbb{W}^{\dag}\mathbb{W}_{t}e^{-i\tilde{H}t/\hbar}
	(\rho(0)\otimes\rho_{T})e^{i\tilde{H}t/\hbar}\mathbb{W}^{\dag}_{t}\mathbb{W}
	\right].
	$
	Here $\rho(0)=|\psi\rangle\langle \psi|$ denotes the initial density 
	matrix of the two qubits.
	The phonon induced evolution results in pure dephasing, meaning that
	the diagonal elements of the DQD density matrix remain constant.
	Using the rules for multiplying and averaging Weyl operators \cite{roszak06b}
	one finds  
	the evolution of the off-diagonal elements of the DQD density matrix,
	\begin{equation}\label{mat-elem}
	{\rho}(t)_{ij}=\langle i|\rho(t)|j\rangle = \rho_{ij}(0)
	e^{i(E_j-E_i)t/\hbar}e^{-iA_{ij}(t)+B_{ij}(t)},
	\end{equation}
	with $E_0=0$, $E_1=E_L$, $E_1=E_R$, and $E_1=E_L+E_R$.
	For two QDs on the same plane (not on top of each other) 
	the decoherence 
	is governed by the functions
	\begin{subequations}
		\label{mat-ev}
		\begin{eqnarray}
		A_{01}&=&A_{02} = \sum|\gk|^{2}\sin\wk t,\\
		A_{03} &=& 4\sum|\gk|^{2}\cos^{2}\bigg(\frac{k_{x}d}{2}\bigg)\sin\wk t,\\
		A_{12}&=&0,\\
		A_{13}&=&A_{23}=A_{03}-A_{01},\\
		B_{01}&=&B_{02}=B_{13}=B_{23}=\sum|\gk|^{2}(\cos\wk t-1)(2n_{\bm{k}}+1),\\
		B_{03}&=&4\sum|\gk|^{2}\cos^{2}\bigg(\frac{k_{x}d}{2}\bigg)(\cos\wk t-1)(2n_{\bm{k}}+1),\\
		B_{12}&=&  4B_{01}-B_{03}=
		4\sum|\gk|^{2}\sin^{2}\bigg(\frac{k_{x}d}{2}\bigg)(\cos\wk
		t-1)(2n_{\bm{k}}+1).
		\end{eqnarray}
	\end{subequations}
	Here $\gk=\fk/\hbar\wk$
	and the individual dot coupling constants were taken to be
	$\fk^{(1,2)}=\fk e^{\pm ik_{x}d/2}$, meaning that the exciton-phonon coupling
	is the same for each dot and they are displaced by the distance $d$
	along the $x$ axis (in-plane).
	For long times, the factors $\cos\wk t$ become quickly oscillating
	functions of $\bm{k}$ and their contribution averages to 0.
	Consequently, $B_{ij}$ decrease form their initial
	value of 0 to a certain asymptotic value depending on the material
	parameters, system geometry and temperature.
	For large distances between the dots $d$, on the other hand, 
	the distance-dependent
	factors $\cos^{2}(k_{x}d/{2})$ and
	$\sin^{2}(k_{x}d/{2})$ become quickly oscillating 
	and their contribution averages to $1/2$
	which leads to $B_{03}=B_{12}=2B_{01}$ and $A_{03}=2A_{01}$.
	In this case the evolution of each qubit can be described separately
	and the single qubit coherence decays following Eq.~(\ref{mat-elem})
	with $A_{01}$ and $B_{01}$,
	which concurs with the fact that distant QDs interact with practically
	separate environments.
	
	The results shown in the previous sections
	have been obtained using the following parameters.
	The exciton wave functions have been modeled by anisotropic Gaussians
	with the extension $l_{\mathrm{e/h}}$ in the $xy$ plane
	for the electron/hole, and $l_{z}$
	along $z$ for both particles.
	The coupling constants for the deformation potential coupling between
	confined charges and longitudinal phonon modes then have the form
	$
	\fk=\sqrt{\frac{\hbar k}{2\varrho Vc}}e^{-l_{z}^{2}k_{z}^{2}/4}
	\left[\sigma_{\mathrm{e}}e^{-l_{\mathrm{e}}^{2}k_{\bot}^{2}/4}
	-\sigma_{\mathrm{h}} e^{-l_{\mathrm{h}}^{2}k_{\bot}^{2}/4} \right].
	$
	Here
	$V$ is the normalization volume of the bosonic environment, 
	$k_{\bot},z$ are phonon momentum
	components in the $xy$ plane and along the $z$ axis,
	$\sigma_{\mathrm{e,h}}$ are deformation potential constants for
	electrons and holes respectively, $c$ is the speed of longitudinal sound,
	and $\varrho$ is the crystal density.
	In our calculations we put $\sigma_{\mathrm{e}}=8$ eV,
	$\sigma_{\mathrm{h}}=-1$ eV, $c=5.6$ nm/ps, $\varrho=5600$ kg/m$^{3}$
	(corresponding to GaAs), and
	$l_{\mathrm{e}}=4.4$ nm, $l_{\mathrm{h}}=3.6$ nm, $l_{z}=1$ nm.
	These parameters correspond to small self-assembled QDs \cite{kroutvar04,greilich06b}.
	
	\subsection*{Rescaled Discord}
	To measure the quantum correlations generated between the two qubits
	we used a geometric measure of the quantum discord called
	the rescaled discord \cite{tufarelli13}. The rescaled discord is 
	related to the geometric discord \cite{dakic10},
	which is defined as the Hilbert-Schmidt distance between
	a given state and the nearest zero-discord state, but it is independent
	of the purity of the studied state.
	For two qubits the relation between the discord measures is
	\begin{equation}
	\label{rd}
	D(\rho)=\frac{1}{2}\left( 1-\frac{\sqrt{3}}{2}\right)^{-1}\left[
	1-\sqrt{1-\frac{D_S(\rho)}{2\Tr\rho^2}}
	\right],
	\end{equation}
	where $D$ denotes the rescaled discord, $D_S$ the geometric discord
	and $\Tr\rho^2$ is the purity of the two qubit density matrix.
	
	For two qubits, it is possible to find the lower and upper bounds
	of the geometric discord (and consequently the rescaled discord)
	given the density matrix of the state.
	The lower bound on the geometric discord
	is given by \cite{dakic10}
	$
	\label{lower}
	D_S'=\frac{1}{4}\max\left(
	\Tr[K_x]-k_x,\Tr[K_y]-k_y
	\right),$
	where $k_x$ is the maximum eigenvalue of the matrix $K_x=|x\rangle\langle x|+TT^T$
	and $k_y$ is the maximum eigenvalue of the matrix $K_y=|y\rangle\langle y|+T^T T$.
	Here, $|x\rangle$ and $|y\rangle$
	denote local Bloch vectors with components $x_i=\Tr[\rho_{AB}(\sigma_i\otimes\mathbb{I})]$
	and $y_i=\Tr[\rho_{AB}(\mathbb{I}\otimes\sigma_i)]$, and the elements of the correlation
	matrix $T$ are given by $T_{i,j}=\Tr[\rho_{AB}(\sigma_i\otimes\sigma_j)]$ (stemming from the
	standard Bloch representation of a two-qubit density matrix $\rho_{AB}$).
	The upper bound is given by \cite{miranowicz12}
	$
	\label{upper}
	D_S''=\frac{1}{4}\min\left(
	\Tr[K_x]-k_x+\Tr[L_y]-l_y,\Tr[K_y]-k_y+\Tr[L_x]-l_x
	\right),
	$
	where $l_x$ and $l_y$ are the maximal eigenvalues of the matrices 
	$L_x=|x\rangle\langle x|+T|\hat{k}_y\rangle\langle \hat{k}_y|T^T$ and
	$L_y=|y\rangle\langle y|+T^T|\hat{k}_x\rangle\langle \hat{k}_x|T$, respectively,
	while $|\hat{k}_x\rangle$ and $|\hat{k}_y\rangle$ are the normalized eigenvectors
	corresponding to the eigenvalue $k_x$ of matrix $K_x$ and $k_y$ of matrix $K_y$.
	The final step in acquiring the upper and lower bounds on the rescaled
	discord is 
	inserting the geometric discord values into Eq. (\ref{rd}).
	For two qubits the rescaled discord can vary between zero and one half,
	where $0$ indicates no quantum correlations present between qubits
	and $1/2$ is reserved for maximally entangled states (the biggest possible
	quantum correlations).

	{\noindent\large\bf Acknowledgements\\}
	This work was financially supported by the
	Polish NCN Grant No.~2012/05/B/ST3/02875.
	The authors are grateful to dr \L ukasz Cywi\'nski
	for useful discussions.
	\newline\newline
	{\noindent\large\bf Author contributions\\}
	The initial idea was devised by K.R. The calculations and numerical
	analysis were performed by J.K.
	The manuscript was written mainly by K.R. with contributions from J.K., who also
	prepared the figures.
	\newline\newline
	{\noindent\large\bf Additional information\\}
	{\bf Competing financial interests:} The authors declare that they have no competing financial interests.

\end{document}